 \documentclass[11pt,aps,showpacs,amsmath,dvips,showkeys,preprint,11pt,nofootinbib]{revtex4}

\usepackage{graphicx}   
\usepackage{color}      
\usepackage{braket}
\usepackage[utf8x]{inputenc} 
\def \be  {\begin{equation}}
\def \ee  {\end{equation}}
\def \ee  {\end{equation}}
\def \bea {\begin{eqnarray}}
\def \eea {\end{eqnarray}}

\newcommand{\nn}{\nonumber}

\begin{document}
  
\title{SU($3$) Polyakov Linear-Sigma Model With Finite Isospin Asymmetry: QCD Phase Diagram}

\author{Abdel Nasser Tawfik} 
\email{tawfik@itp.uni-frankfurt.de}
\affiliation{\scriptsize Nile University, Egyptian Center for Theoretical Physics (ECTP), Juhayna Square off 26th-July-Corridor, 12588 Giza, Egypt}
\affiliation{\scriptsize Goethe University, Institute for Theoretical Physics (ITP), Max-von-Laue-Str. 1, D-60438 Frankfurt am Main, Germany}

\author{Abdel Magied Diab}
\email{a.diab@eng.mti.edu.eg}
\affiliation{\scriptsize Modern University for Technology and Information (MTI), 11571 Cairo, Egypt}
\affiliation{\scriptsize World Laboratory for Cosmology And Particle Physics (WLCAPP), 11571 Cairo, Egypt}

\author{M.T. Ghoneim}
\affiliation{\scriptsize Physics Department, Faculty of Science, Cairo University, 12613 Giza, Egypt}

\author{H. Anwer}
\affiliation{\scriptsize Physics Department, Faculty of Science, Cairo University, 12613 Giza, Egypt}
\affiliation{\scriptsize Physics Department, Zewail City of Science and Technology, 12578  Giza, Egypt}

\date{\today}

\begin{abstract}

The SU($3$) Polyakov linear-sigma model (PLSM) in mean-field approximation is utilized in analyzing the chiral condensates $\sigma_u$, $\sigma_d$, $\sigma_s$ and the deconfinement order parameters $\phi$, $\bar{\phi}$, at finite isospin asymmetry. The bulk thermodynamics including pressure density, interaction measure, susceptibility, and second-order correlations with baryon, strange and electric charge quantum numbers are studied in thermal and dense medium. The PLSM results are confronted to the available lattice QCD calculations. The excellent agreement obtained strengthens the reliability of fixing the PLSM parameters and therefore supports further predictions even beyond the scope of the lattice QCD numerical applicability. From the QCD phase structure at finite isospin chemical potential ($\mu_I$), a novel expression for the explicit symmetry breaking term $h_3$ is introduced,
we find that the pseudo-critical temperatures decrease with the increase in $\mu_I$. We conclude that the QCD phase structure in ($T_\chi$-$\mu_I$) plane seems to extend the one in ($T_\chi$-$\mu_B$) plane.

\end{abstract}


\maketitle

\tableofcontents
\makeatletter
\let\toc@pre\relax
\let\toc@post\relax
\makeatother

\section{Introduction \label{intro}}

The isobaric quantum spin asymmetry in up ($u$) and down ($d$) quarks introduced by Heisenberg likely gives an plausible explanation for the mass difference between proton and neutron, for instance. This could be expressed as a vector quantity with the 3rd-component having $​1/2$ and $-1/2$, respectively, but entirely vanishing for all other quark flavors. Finite isospin plays a major role in various physical systems, for instance, early Universe especially at large lepton asymmetry, compact stars with pion condensates, and spectroscopy of nuclei \cite{Wigner:1936dx}. 

While in lattice QCD simulations at finite baryon chemical potential ($\mu_B$) a complex action appears, known as sign problem, fortunately finite isospin chemical potential ($\mu_I$) has a real and positive action and therefore can straightforwardly be implemented in the Monte Carlo (MC) techniques. Hence, QCD at finite $\mu_I$ could be utilized to test the various methods attempting to overcome the sign problem in lattice QCD simulations at finite $\mu_B$, such as Taylor expansion \cite{Philipsen:2007rj}. Furthermore, QCD at finite $\mu_I$ \cite{Son:2000xc} and QCD at finite $\mu_B$ \cite{Philipsen:2007rj} obviously share some common features, such as deconfinement, particle creation, Silverblaze phenomenon, and Bose-Einstein condensation (BEC) at large densities. Accordingly, the indirectly drawing of qualitative conclusions on QCD at finite $\mu_B$, studying QCD matter at finite $\mu_I$ gives different interesting features \cite{Brandt:2017zck, Kovacs:2007sy}, for instance, enriching the QCD phase diagram, forming BEC and the yet-still-hypothetical superconducting phases \cite{Detmold:2012wc}. 

The chiral isospin asymmetry is an active research aiming at characterizing the imbalance between the charged pion degrees-of-freedom \cite{Li:1997px} that could be formed in dense quark matter, e.g. neutron stars \cite{Migdal:1990vm,Steiner:2004fi}. To this end, the relevant variables of QCD matter are temperatures ($T$) and number densities ($n_f$ for $f$-th quark flavors). In SU$(2)$, the baryonic ($n_B$) and isospin number density ($n_I$) related to the light sector of quark flavors [up ($u$) and down ($d$)] such as  $n_B=(n_u+n_d)/3$ and $n_I=n_u-n_d$, respectively. 

At the experimental site, the LHCb collaboration has recently analyzed the decays and the partial branching ratios of neutral and charged boson (B) as functions of the dimuon mass squared and found that the isospin asymmetries are consistent with the Standard Model, while the measured branching ratios are smaller than the respective theoretical predictions \cite{Aaij:2014pli}. The isospin asymmetry enhances the $\eta$, $\eta^{\prime}$, $\pi^0$ mixing, which likely comes up with an additional contribution to the amplitude of the decay process $B\rightarrow \pi\pi$. Furthermore, understanding the properties of matter at high density, which are theoretically very challenging and experimentally still not well accessible, such as the transition to hyperonic matter, BEC, and color super conductivity (CSC), is essential for the stellar properties of neutron star  (NS). Reliable equations of state (EoS), as the ones intended to be deduced from the chiral quark model, the Polyakov linear-sigma model (PLSM), would make it possible to analyze the impacts of the isospin asymmetry on the early Universe, high-density matter, and NS binary mergers. 

It was pointed out that the asymmetry between $u$- and $d$-quark affects the phase structure of the QCD matter \cite{Khunjua:2017mkc}. The possible medium effects due to finite isospin asymmetry, such as modifications in the energies of kaons and antikaons \cite{Mishra:2008dj}, are crucial for asymmetric heavy-ion collisions, especially, the neutron-rich ones at the future facilities NICA and FAIR. These possible modifications, for example, lead to a decrease in the antikaon mass, which could be understood due to interactions with nucleons and scalar mesons. It is apparent that these are also essential for NS phenomenology at finite $\mu_B$ and finite $\mu_I$ \cite{Steiner:2004fi}. The recent gravitational-wave observations of NS binary mergers open new research directions not only in cosmology and astrophysics but concretely in proposing EoS for NS and describing how this looks like in the post-merger ring-down phase, that allows shaping deformed, oscillating, differentialy rotating, and very massive NS \cite{Takami:2015gxa}. 

With highlighting some studies conducted in the PLSM for the characterization of the thermal QCD phase structure at vanishing and finite baryon density \cite{AbdelAalDiab:2018hrx,Tawfik:2016gye,Tawfik:2015tga,Tawfik:2014gga,AbdelAalDiab:2018hrx}, the intension of extending this chiral model to finite isospin asymmetry carried out in the present script can be well endorsed. The SU($3$) PLSM was utilized in analyzing higher-order moments of the particle multiplicity \cite{Tawfik:2014gga} and in characterizing the temperature dependence of the transport and conductivity coefficients and was compared with recent lattice QCD calculations \cite{Tawfik:2016edq} at finite magnetic fields \cite{Tawfik:2016ihn}. The inclusion of charm quark was also proposed, see refs. \cite{AbdelAalDiab:2018hrx,Diab:2016iig}.

The present script is organized as follows. The SU($3$) Polyakov linear-sigma model (PLSM) is introduced in Sect. \ref{eLSM}. The essential expressions at finite isospin asymmetry are outlined in Sect. \ref{model}. In Sect. \ref{resulat}, we investigate the impacts of the isospin asymmetry on the QCD phase transition(s). The PLSM order parameters are discussed in Sect. \ref{Orders}. In Sect. \ref{Thermo}, we discuss on the resulting PLSM thermodynamics as functions of temperature and finite isospin chemical potential. We estimate the fluctuations of the conserved charges in Sect. \ref{Fluc}. Finally, we introduce the chiral phase transition as pseudo-critical temperature at finite isospin chemical potential in Sect. \ref{QCD}. Last but not least, Sect. \ref{conclusion} is devoted to the conclusions.

\section{SU($3$) Polyakov Linear-Sigma Model \label{eLSM}}

The theory of strong interactions, the quantun chromodynamcs (QCD), in thermal and dense medium plays a crucial role in explaining a wide range of physical phenomena. Study of the chiral phase structure sheds light on the evolution of the high-energy collisions, the interior structure of the stellar compact objects, and physics of early Universe. Besides heavy-ion experiments such as Large Hadron Collider (LHC) at CERN and the Relativistic Heavy Ion Collider (RHIC) at BNL, the lattice QCD simulations help in exploring the phase structure of the QCD matter at vanishing and finite baryon density \cite{Bernard:2004je,Hands:2001ee, Kogut:2001na, Kogut:2001if,Kogut:2002kj,Aoki:2006we, Alles:2006ea, Hands:2006ve, Hands:2010gd, Borsanyi:2011sw, Borsanyi:2016ksw, Bazavov:2011nk, Bazavov:2014pvz, Karsch:2013fga, Aoki:2009sc}.

The limitations of MC techniques at finite baryon chemical potential could be seen as promoters for unavoidable utilization of the various QCD-like approaches which are relaibly able to explain a wide range of QCD phenomena such as the bulk properties and the thermodynamic fluctuations of the conserved charges, and the chiral quark-hadron phase transitions, as well \cite{Asakawa:1989bq,Fukushima:2008wg,Ratti:2005jh,Carignano:2010ac,Bratovic:2012qs,Rischke:2003mt,Schaefer:2006ds,Schaefer:2007pw,Schaefer:2008ax,Schaefer:2009st}.

The present study aims at analyzing the impacts of the finite isospin asymmetry in the chiral models, such as PLSM, which are particularly helpful in characterizing the thermodynamic properties of the QCD phase structure. Concretely, it intends to distinguish between the light quarks in thermal and dense medium and to confront the PLSM results to recent lattice QCD calculations. Moreover, a general expression of the chiral limit at finite isospin asymmetry shall be proposed.  

\subsection{PLSM formalism at finite isospin Asymmetry \label{model}}

In Minkowski space,  the LSM Lagrangian density with $N_f$ quark flavors can be incorporated with the Polyakov-loop potential
\bea
\mathcal{L}_{PLSM} = \mathcal{L}_{\mathrm{chiral}}-\mathbf{\mathcal{U}}(\phi, \bar{\phi}, T). \label{PLSM_Eq}
\eea
\begin{itemize}
\item The first term in rhs of Eq. (\ref{PLSM_Eq}) stands for the LSM Lagrangian density in the chiral limit and is given as 
\bea
\mathcal{L}_{\mathrm{chiral}} &=&\mathcal{L}_{\overline{\psi} \psi}+\mathcal{L}_m, 
\eea
where the first term counts for the contributions of the quarks (fermions) with $N_c$ color degrees-of-freedom while the second term stands for the mesonic (bosonic) fields. 
\item The second term in rhs of Eq. (\ref{PLSM_Eq}), $\mathcal{U}(\phi, \bar{\phi},T)$ stands for the Polyakov-loop potential, which introduces the gluonic degrees-of-freedom and the dynamics of the quark-gluon interactions to the chiral LSM \cite{Fukushima:2008wg}. In the present calculations, we utilize a Polyakov-loop potential, which counts for strong coupling and includes higher orders of the Polyakov-loop variables.
\end{itemize}
These are assumed to characterize the QCD symmetries in pure-gauge theory \cite{Ratti:2005jh,Roessner:2006xn,Schaefer:2007pw,Fukushima:2008wg, Lo:2013hla}
\bea
\mathcal{L}_{\overline{\psi} \psi} &=& \sum_f \overline{\psi}_f(i\gamma^{\mu} D_{\mu}-g\,T_a(\sigma_a + i \gamma_5 \pi_a))\psi_f, \label{lfermion}  \\ 
\mathcal{L}_m &=&\mathrm{Tr}(\partial_{\nu} \Phi^{\dag}\partial^{\nu} \Phi-m^2
\Phi^{\dag} \Phi)-\lambda_1\,[\mathrm{Tr}\,(\Phi^{\dag} \Phi)]^2 \nonumber \\
&-& \lambda_2\,\mathrm{Tr}(\Phi^{\dag} \Phi)^2+c[\mathrm{Det}(\Phi)+\mathrm{Det}(\Phi^{\dag})]
+\mathrm{Tr}[H(\Phi+\Phi^{\dag})],  \hspace*{6mm} \label{lmeson} \\ 
 \mathbf{\mathcal{U}}_{\mathrm{Fuku}}(\phi, \bar{\phi}, T) & = & - b \;T \left[54\, \phi \,\bar{\phi} \;\exp(-a/T) + \ln(1-6 \phi \bar{\phi} - 3(\phi \bar{\phi})^2+4 (\phi^3 + \bar{\phi}^3))\right]. \label{FUKU}
\eea
where $D_{\mu},\; \mu,\; \gamma^{\mu}$ and $g$  are covariant derivative, Lorentz index, chiral spinors, and Yukawa coupling constant, respectively. $\psi$ are a Dirac spinor fields for the quark flavors $f=[u,\, d,\, s]$. The explicitly symmetry breaking, $H= \hat{T}_a h_a$, where $h_a$ is a nine parameters of the explicitly symmetry breaking in SU($3$). As a result, the diagonal components of the symmetry generators $h_0,\, h_3,\, h_8$ are non-vanishing. Moreover, the mesonic field $\Phi$ is a $(3\times3)$ matrix for nonet meson states,
\bea
 \bar{\Phi} &=&  \sum_{a=0}^{N_f^2 - 1} T_a ( \bar{\sigma_a}+i  \bar{\pi_a})  \label{mesonfield}
\eea
where $\sigma_a$ and  $\pi_a$ are the scalar and pseudoscalar fields, respectively. In vacuum state with $U(1)_A$ anomaly and as a result of the spontaneous symmetry breaking, the  expectation values of mesonic fields, $\braket{\Phi}$, and of their conjugates, $\braket{\Phi^{\dag}}$ are generated with the quantum numbers of the vacuum \cite{Gasiorowicz:1969kn}. This leads to exact vanishing mean value of  $\bar{\pi}_a$ but assures finite mean value of $\bar{\sigma_a}$ corresponding to the diagonal generators $U(3)$ as $\bar{\sigma_0} \neq\, \bar{\sigma_3}\neq\, \bar{\sigma_8} \neq 0$ , where $\braket{\Phi} =  T_0 \bar{\sigma_0} + T_3 \bar{\sigma_3}+ T_8 \bar{\sigma_8}$ . 

On the other hand, $\bar{\sigma_3}$ breaks the isospin asymmetry SU($2$) \cite{Gasiorowicz:1969kn}, Furthermore, the potential of pure mesonic contributions in SU($N_f$) can be written as \cite{Lenaghan:2000ey},
\bea
U(\bar{\sigma}) &=& \left(\frac{ m^2}{2} -h_a\right) \bar{\sigma}_a - 3 \mathcal{G}_{abc}    \bar{\sigma}_b  \; \bar{\sigma}_c  - \frac{4}{3} \mathcal{F}_{abcd}\; \bar{\sigma}_b  \; \bar{\sigma}_c  \bar{\sigma}_d, \label{Usigma_a}
\eea
where the coefficeints $\mathcal{G}_{abc}$ and $\mathcal{F}_{abcd}$ are given as \cite{Lenaghan:2000ey}
\bea
\mathcal{G}_{abc} &=&  \frac{c}{6} \left[ d_{abc}  - \frac{3}{2} \left(  d_{0bc} \delta_{a0} + d_{a0c}  \delta_{b0}   + d_{ab0}  \delta_{c0}  \right) +\frac{9}{2} d_{000} \delta_{a0} \delta_{b0} \delta_{c0}\right], \\ 
\mathcal{F}_{abcd} &=&  \frac{\lambda_1}{4}\left[ \delta_{ab} \delta_{cd}  + \delta_{ad} \delta_{cd} + \delta_{ac} \delta_{bd}\right] + \frac{\lambda_2}{8}\left [d_{abn} d_{ncd} +  d_{adn} d_{nbc} +d_{acn} d_{nbd} \right].
\eea
The explicitly symmetry breaking terms, $h_0, h_3$ and $h_8$, can be determined by minimizing the potential, Eq. (\ref{Usigma_a}), on tree level, $\partial U(\bar{\sigma})/ \partial \bar{\sigma}_a =0$. $h_0$ and $h_8$, can be determined from the partially conserved axial current (PCAC) relations (see App.  [\ref{App_algebra})] \cite{Lenaghan:2000ey} 
\bea
h_0 &=& \frac{1}{\sqrt{6}}\left( m_\pi^2 f_\pi + 2 m_K^2 f_K \right), \\  
h_8 &=& \frac{2}{\sqrt{3}}\left( m_\pi^2 f_\pi - m_K^2 f_K \right).
\eea
Thus, the explicit symmetry breaking term, $h_3$, can be deduced from $\partial U(\bar{\sigma})/ \partial \bar{\sigma}_3  =0$,  
\bea
h_3 &=& \left[ m^2 + \frac{c}{\sqrt{6}} \bar{\sigma_0} - \frac{c}{\sqrt{3}} \bar{\sigma_8} + \lambda_1 \left( \bar{\sigma_0}^2 + \bar{\sigma_3}^2 + \bar{\sigma_8}^2 \right) +   \lambda_2   \left(   \bar{\sigma_0}^2 +   \frac{ \bar{\sigma_3}^2}{2} +   \frac{ \bar{\sigma_8}^2}{2}+  \sqrt{2} \bar{\sigma_0}  \bar{\sigma_8} \right) \right] \bar{\sigma_3}, 
\eea
where the square brackets $[\cdots]$ is the squared mass of the $a_0$ meson and $\bar{\sigma}_3 =\left(f_{K^{\pm}}-f_{K^{0}} \right)$  
\bea
h_3 &=&  m^2_{a_0} \left(f_{K^{\pm}}-f_{K^{0}} \right), 
\eea
As a result of the finite isospin asymmetry, the masses of the quark flavors, as nature likely prefers, are not entirely degenerated, i.e. $m_u \neq m_d \neq m_s$. To assure this situation, we use the orthogonal basis transformation to convert the condensates from the original basis, $\sigma_0,\; \sigma_3$, and $\sigma_8$ to pure up ($\sigma_u$), down ($\sigma_d$), and strange ($\sigma_s$) quark flavor basis, respectively, 
\bea  
\begin{bmatrix}
     \bar{\sigma_u} \\    \bar{\sigma_d} \\  \bar{\sigma_s}    
\end{bmatrix} = \frac{1}{\sqrt{3}} 
\begin{bmatrix}
      \sqrt{2} &1 & 1 \\
      \sqrt{2} &-1 & 1 \\
      1 & 0& -\sqrt{2} \\
\end{bmatrix} 
\begin{bmatrix}
    \bar{\sigma_0} \\    \bar{\sigma_3} \\  \bar{\sigma_8}   
\end{bmatrix}. \label{eq:sigmas}
\eea
Accordingly, the masses of $u$, $d$, and $s$ quarks can be expressed as, 
\bea
m_u = \frac{g}{2} \sigma_u, \quad  \quad m_d = \frac{g}{2} \sigma_d, \quad \quad m_s = \frac{g}{\sqrt{2}} \sigma_s.  \label{chiral_mass}
\eea
As mentioned above, the potential of the pure mesonic contributions can be obtained by substituting the mesonic field, Eq. (\ref{mesonfield}), in the potential term of chiral LSM Lagrangian density, Eq. (\ref{Usigma_a}). The potential of mesonic contributions can be given as,
\bea
U(\sigma_u,\, \sigma_d,\, \sigma_s) &=& \frac{m^2}{4}\Big[\sigma_u^2 +\sigma_d^2 + 2 \sigma_s^2 \Big] - \frac{c}{2\sqrt{2}} \sigma_u\, \sigma_d \, \sigma_s + \frac{\lambda_1}{16} \Big( \sigma_u^2 + \sigma_d^2  +2\sigma_s^2 \Big)^2  \nn \\ &+& \frac{\lambda_2}{16} \Big(\sigma_u^4 + \sigma_d^4  +4 \sigma_s^4  \Big) - h_{ud} \frac{\sigma_u+\sigma_d}{2} - h_3 \frac{\sigma_u -\sigma_d}{2} - h_s \sigma_s.  \label{pureLSMPOT} 
\eea   
For symmetry breaking in vacuum, $H\neq 0,\; c\neq 0$ and $\lambda\neq 0$, the impacts of the isospin asymmetry violating SU($2$), $h_0,\; h_3$ and $h_8$, have nonzero values, at the chiral masses $m_u \ne m_d \ne m_s \ne 0 $. 

The LSM parameters, $m^2$, $h_l,\; h_s, \; h_3,\; \lambda_1,\; \lambda_2$, and $c$ can be given in dependence on $m_\sigma$ \cite{Schaefer:2008hk}. Tab. \ref{tab:1a} summarizes these parameters, at $m_\sigma=800~$MeV \cite{Schaefer:2008hk}. It should be noticed that the isospin parameters, $\sigma_3$ and $h_3$, are greatly differentiate between the SU$(2)$ quark flavors, as well.

\begin{table}[htb]
\begin{center}
\begin{tabular}{|c | c | c | c | c | c | c | c |}
\hline
$m_\sigma$ [MeV] & $c\,$ [MeV] & $h_{ud}\,$ [MeV$^3$]& $h_3\,$ [MeV$^3$] & $h_s\,$ [MeV$^3$] & $m^2 \,$ [MeV$^2$] & $\lambda _1$ & $\lambda _2$\\ 
\hline 
$800$ & $4807.84$ & $(120.73)^3$ & $-(78.31)^3$ & $(336.41)^3$ & -$(306.26)^2$ & $13.49$& $46.48$\\ 
\hline 
\end{tabular}
\caption{Values of the LSM parameters given in the mesonic Lagrangian, Eq. (\ref{lmeson}), as fixed at $m_\sigma=800~$MeV \cite{Schaefer:2008hk}.}  \label{tab:1a}
\end{center}
\end{table} 

In the mean-field approximation (MFA), the PLSM thermodynamic potential can  be related the to grand-canonical function  $\mathcal{Z}$, which is given in dependence of the temperatures $T$ and the chemical potentials of $f-$th quark flavor $\mu_f$, see App. (\ref{App_A}),   
\bea
\Omega(T, \mu_f) = \frac{-T\;\cdot \ln{\left[\mathcal{Z}\right]}}{V} 
&=& U(\sigma_u,\, \sigma_d,\, \sigma_s)+\mathbf{\mathcal{U}}_{\mathrm{Fuku}}(\phi, \bar{\phi}, T)+\Omega_{\bar{\psi}\psi} (T,\,\mu_f). \label{potential}
\eea  
The chemical potentials $\mu_f$  are related to conserved quantum numbers of - for instance - baryon number ($B$), strangeness ($S$), electric charge ($Q$), and isospin ($I$) of each quark flavors, 
\bea
\mu_u &=& \frac{\mu_B}{3} + \frac{2 \mu_Q}{3} + \frac{\mu_I}{2}, \label{muf-eqs1}\\
\mu_d &=& \frac{\mu_B}{3} - \frac{\mu_Q}{3} - \frac{\mu_I}{2},  \label{muf-eqs2} \\
\mu_s &=& \frac{\mu_B}{3} - \frac{\mu_Q}{3} -\mu_S. \label{muf-eqs3}
\eea
In expression (\ref{potential}), the first term $U(\sigma_u,\, \sigma_d,\, \sigma_s)$; the potential of the pure mesonic contributions, was given Eq. (\ref{pureLSMPOT}), while the second term $ \mathbf{\mathcal{U}}_{\mathrm{Fuku}}(\phi, \bar{\phi}, T)$, the potential of Polyakov loop variables, was elaborated in Eq. (\ref{FUKU}). The last term refers to the quarks and antiquarks contributions to the PLSM potential \cite{Fukushima:2008wg,Kapusta:2006book,Mao:2009aq,Schaefer:2009ui},
\bea
\Omega_{\bar{\psi}\psi}(T, \mu _f)&=& -2 \,T \sum_{f=u, d, s} \int_0^{\infty} \frac{d^3\vec{P}}{(2 \pi)^3}   \ln \left[ 1+n_{q,f}(T, \; \mu_f) \right]   +\ln \left[ 1+ n_{\bar{q},f}(T, \; \mu_f) \right],  \label{PloykovPLSM}  \eea 
where the number density distribution for particle is given as
\bea 
n_{q,f}(T, \; \mu_f)&=& 3\left(\phi+\bar{\phi}e^{-\frac{E_f-\mu _f}{T}}\right)\times e^{-\frac{E_f-\mu _f}{T}}+e^{-3 \frac{E_f-\mu _f}{T}},
%
\eea
which is identical to that of anti-particle $n_{q,f}(T,\; \mu_f)$ with $-\mu_f$ replacing $+\mu_f$ and the order parameter  $\phi$ by its conjugate $\bar{\phi}$ or vice versa. $E_f=(\vec{P}^2+m_f^2)^{1/2}$ is the energy-momentum dispersion relation with $m_f$ being the mass of $f^{th}$ quark flavor. 

With this regard, we extend PLSM towards analyzing the QCD phase structure and the thermodynamic properties to finite isospin asymmetry. Firstly, the effects of finite isospin asymmetry on differentiation between the nonstrange condensates of $u$- and $d$-quark shall be analyzed. Secondly, as a result of the isospin symmetry breaking, $\sigma_3$ should have a nonzero value because $\sigma_u=\sigma_l+\sigma_3$ and $\sigma_d=\sigma_l-\sigma_3$. To this end, we estimate the pure mesonic potential for $N_f$ quark flavors, Eq. (\ref{pureLSMPOT}), as functions of temperatures and chemical potentials. It can be noticed that the solutions of the gap equations, Eqs. (\ref{Gap-eq}), at finite saddle point in vacuum, i.e. $T=\mu_f=0$, and Yukawa coupling constant $g=6.5$,  result in expectation values of the nonstrange chiral condensates $\sigma_{u0}= 91.94~$MeV, $\sigma_{d0}= 92.94~$MeV and $\sigma_{l0}= 92.4~$MeV, while the strange chiral condensate is $\sigma_{s0}= 94.5~$MeV. The importance of this finding is that it gives predictions for the nonstrange and strange quark mass, $m_l \approx 300~$MeV and $m_s \approx 434.5$, respectively. Furthermore, there are various physical quantities characterizing the QCD phase structure in thermal and dense medium can straightforwardly be deduced.

\section{Results and discussion \label{resulat}} 

For a reliable differentiation between $u$- and $d$-quark condensates, we need to estimate the influences of finite isospin on the PLSM chiral condensates and the deconfinement order parameters. Then, we calculate the thermal behavior of the conserved charge fluctuations at vanishing and finite isospin chemical potential $\mu_I$.  Last but not least, we introduce the variation of the pseudo-critical temperatures with the normalized isospin chemical potentials.

\subsection{PLSM condensates and order parameters \label{Orders}}

In this section, we evaluate the condensates $\sigma_u$, $\sigma_d$ and $\sigma_s$ and the Polyakov-loop fields, $\phi$ and $\bar{\phi}$, known as chiral and deconfinement order parameters, respectively, in mean-field approximation. To this end, we start with the real part of the thermodynamic potential $\mathcal{R}e\;[\Omega(T,\; \mu_f)]$, Eq. (\ref{potential}). This potential part should be minimized at a saddle point. With the solutions of the gap equations, Eq. (\ref{Gap-eq}); a complete set of equations, one can analyze the behaviors of $\sigma_u$, $\sigma_d$, $\sigma_s$, $\phi$, and $\bar{\phi}$ in thermal and dense medium. By solving the gap equations, one could recognize that the thermodynamic potential mainly depends on two independent variables; the temperature $T$ and the chemical potential $\mu_f$. The latter is related to the quark flavors $f$ as expressed in Eq. (\ref{muf-eqs}).

\begin{figure}[htb]
\centering{
\includegraphics[width=16.cm, height=11.cm,angle=0]{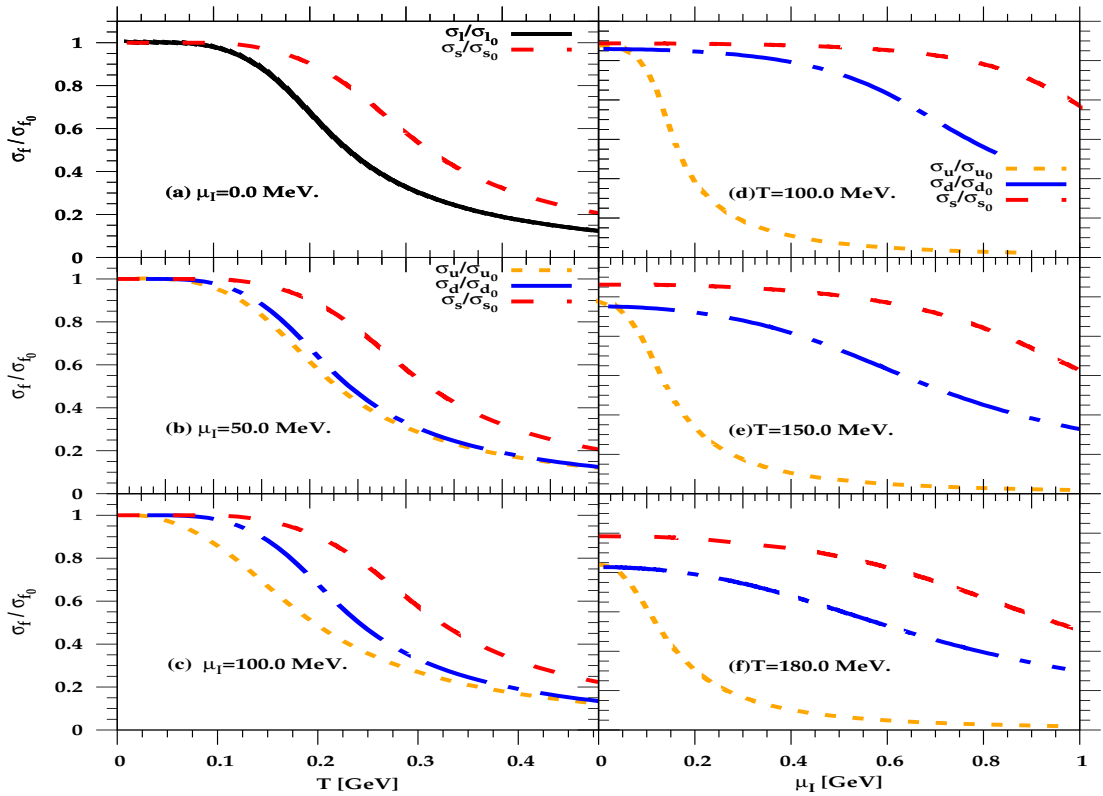}
\caption{\footnotesize (Color online) Top-left panel (a) depicts the temperature dependence of the normalized condensates for nonstrange $\sigma_l/\sigma_{l0}$ (dashed) and strange $\sigma_s/\sigma_{s0}$ quarks (solid curve) at $\mu_I =0.0$. Middle and bottom panels (b) and (c) illustrate the same as (a) but at $\mu_I=50.0$ and $100.0~$MeV, respectively. $\sigma_u/\sigma_{u0}$ and $\sigma_d/\sigma_{d0}$ are shown as dotted and dash-dotted curves, respectively. Right panels present  the same as the left panels but as functions of $\mu_I$ at $T=100.0$ (d), $150.0$ (e) and $180.0~$MeV (f).  
\label{fig_crossover}}    
}
\end{figure}

Figure \ref{fig_crossover} depicts the temperature dependence of the normalized chiral condensates $\sigma_f/\sigma_{f0}$ in thermal (left panel) and dense medium (right panel), from which the impacts of $\mu_I$ on the quark-hadron phase transitions could be estimated, at least qualitatively. The left panels depict the temperature dependence at (a) $\mu_I=0.0$, (b) $50.0$, and (c) $100.0~$MeV. At vanishing $\mu_I$, the expectation values of the nonstrange condensates $\sigma_l=(\sigma_u+\sigma_d)/2$ could be determined as $\sigma_l/\sigma_{l0}$ and $\sigma_s/\sigma_{s0}$, at varying $T$. 

In the hadronic phase, i.e. $T<T_\chi$ where $T_\chi$ is the pseudo-critical temperature, the condensates start with the same normalized vacuum value, e.g. $\sigma_{l0}$, and remain almost unchanged, at $T<<T_\chi$. This means that below $T_\chi$ all normalized quark condensates are nearly entirely non-distinguishable. Under these conditions, the system is apparently confined. A further increase in $T$ is accompanied by a slow decrease in the {\it bundled} chiral condensates drawing two differentiable curves characterizing nonstange and strange condensates. The smooth decrease obviously describes a slow transition (crossover). At temperatures larger than $T_\chi$, the three types of chiral condensates are remarkably suppressed and the QCD system is converted into a deconfined state. 

At finite $\mu_I$, i.e. middle and bottom panels (b) and (c), the significance of isospin parameters $\sigma_3$ and $h_3$ apparently comes into play an essential role. This leads to differentiation between the $u$ and $d$ chiral condensates. When $T$ approaches $T_\chi$, the normalized nonstrange condensates split into two different curves, especially within the region of phase transition; dotted and dash-dotted curve, respectively, indicating that $u$ and $d$ chiral condensates become distinguishable. Accordingly, the chiral pseudo-critical temperatures ($T\chi$) can be estimated, at least qualitatively. We notice that $T_\chi$ decreases with increasing $\mu_I$. At higher temperatures, the light quark condensates become more and more suppressed. These results point out to a crossover transition in the ($T\chi$-$\mu_I$) plane, as presented in Fig \ref{QCDdiagram}. Obviously $\sigma_s/\sigma_{s0}$ remains nearly unaffected with the increase in $\mu_I$. The strange quark condensate shows almost the same behavior as in panel (a).

The right panel of Fig. \ref{fig_crossover} shows the $\mu_I$-dependence of $\sigma_f/\sigma_{f0}$ of $f$-th quark flavor at $T=100.0$ (d), $150.0$ (e) and $180.0~$MeV (f). These values are chosen to characterize the hadronic phase. We notice that the qualitative behavior is nearly similar to the left panel. As $T$ increases, a rapid drop takes place but also the three curves proceed, entirely differently.


At $T=100.0~$MeV [panel (d)], we notice that the chiral condensates for $u$- and $s$-quark seem to remain unaffected at very large $\mu_I$, while that of $d$-quark is relatively more affected. This can be detailed as follows. $\sigma_s/\sigma_{s0}$ remains longer than $\sigma_d/\sigma_{d0}$, which in turn is not as sensitive to $\mu_I$ as $\sigma_u/\sigma_{u0}$. This means that the corresponding pseudo-critical temperature, which is an approximately averaged value where the condensate rapidly declines, strongly depends on the quark flavor. We observe that $\sigma_u/\sigma_{u0}$ and $\sigma_d/\sigma_{d0}$ are slightly smaller than unity, at vanishing $\mu_I$, while $\sigma_s/\sigma_{s0}$ starts being affected at $\mu_I\gtrsim 0.5~$GeV. This indicates that finite $\mu_I$ sets on its effects very early, which can also be understood as the vacuum values are apparently altered even at $\mu_I=0$.

As $T$ increases to $150.0~$MeV panel (e), $\sigma_s/\sigma_{s0}$ becomes affected at $\mu_I\gtrsim 0.2~$GeV, while both $\sigma_d/\sigma_{d0}$ and $\sigma_u/\sigma_{u0}$ are influenced at smaller values of $\mu_I$. Also, here finite $T$ seems to alter the vacuum condensates even at vanishing $\mu_I$. Also, we notice that the decrease in $\sigma_u/\sigma_{u0}$ is faster than that of $\sigma_d/\sigma_{d0}$, which is more rapid than $\sigma_s/\sigma_{s0}$. Accordingly, we conclude that the crossover transition becomes slower when moving from $u$- to $d$- and then to $s$-quark flavors \cite{Tawfik:2004vv}.

Similarly, we can analyze the results of the panel (f) as follows. As $T$ approaches $T_\chi$, for instance at $T=180.0~$MeV, the three chiral condensates become slightly smaller than unity, at vanishing $\mu_I$. The conclusion that the averaged value of the chiral condensate rapidly declines and strongly depends on the quark flavors can be drawn, as well.

\begin{figure}[htb]
\centering{
\includegraphics[width=16.cm,height= 7 cm,angle=0]{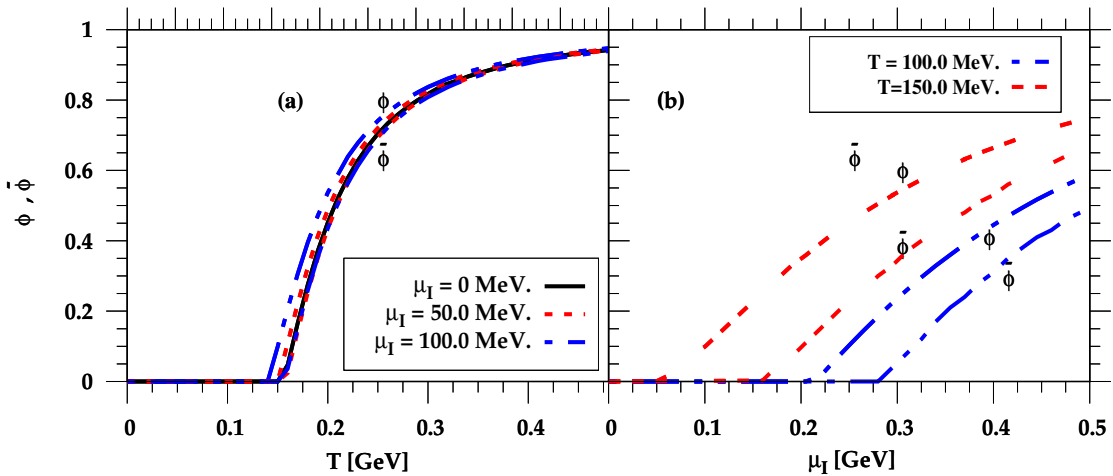}
\caption{\footnotesize (Color online) Left panel (a): the order parameters of the Polyakov-loop variables $\phi$ and $\bar{\phi}$ are given as functions of $T$ at $\mu_I =0.0$ (solid), $50.0$ (dotted), and $100.0~$MeV (dash-double-dotted curves).  Right panel (b): the same as in panel (a) but here in dependence on $\mu_I$ at $T=100.0$ (dashed) and $150.0~$MeV (dash-double-dotted curve).   
\label{Fig.Ploops}}
}
\end{figure}
Figure \ref{Fig.Ploops} presents the order parameters corresponding to the Ployakov-loop variables (related to deconfinement phase transition) $\phi$ and $\bar{\phi}$ in thermal and dense medium as functions of $T$ (left) and of $\mu_I$ (right panel). The left panel shows $\phi$ and $\bar{\phi}$ vs. $T$, at $\mu_I=0.0$ (solid) $50.0$ (dotted), and $100.0~$MeV (dash-double-dotted curve). It is obvious that the thermal evolution of the deconfinement phase transition goes very smooth, i.e. the raise from low to large $\phi$ and $\bar{\phi}$ takes place slowly or within temperatures of couple hundreds MeV. This draws a typical picture about the chiral crossover transition. We also notice that at $\mu_I=0$, the expectation values of $\phi$ and $\bar{\phi}$ are identical, i.e. $\phi=\bar{\phi}$ (solid curves). Increasing $\mu_I$ increases $\phi$, but simultaneously decreases $\bar{\phi}$. Accordingly, $\phi$ and $\bar{\phi}$ become more and more distinguishable when increasing $\mu_I$. The corresponding pseudo-critical temperatures seem very weakly depending on $\mu_I$.
 
The right panel (b) shows the same as in the left panel (a) but here $\phi$ and $\bar{\phi}$ are given in dependence on $\mu_I$ at $T=100.0$ (dash-double-dotted) and $150.0~$MeV (dashed curve). From a large set of calculations at different temperatures (only two values are depicted, here), we conclude that both $\phi(\mu_I)$ and $\bar{\phi}(\mu_I)$ are depending on $T$, as well. It is apparent that $\phi(\mu_I)$ shows a larger increase with $\mu_I$ than $\bar{\phi}(\mu_I)$. It is clear that a pseudo-critical isospin asymmetry could be evaluated from the derivative of $\phi$ and of $\bar{\phi}$ with respect to $\mu_I$. Equivalently, this would be estimated within the $\mu_I$-region, where $\phi$ and $\bar{\phi}$ rapidly increase. Similarly, we could estimated this, where corresponding curves (or their tangents) have the largest slopes. Approximately, we find that the pseudo-critical isospin asymmetry decreases with increasing temperature.
 
\subsection{Bulk thermodynamics \label{Thermo}} 
Various thermodynamic quantities can be estimated from the PSLM thermodynamic potential $\Omega(T,\mu_f)$ or the grand canonical partition function $\mathcal{Z}$, for instance at finite $\mu_f$, we have $p(T,\mu_f) =-\Omega(T,\mu_f)$ and the normalized  interaction measure is given as
\bea  
\frac{\Delta(T,\; \mu_f)}{T^4} &=& T \frac{\partial}{\partial T} \; \Big(\frac{p}{T^4}\Big)=\frac{\epsilon-3p}{T^4},  \label{ThermoEQs} 
\eea 
where $\epsilon= T^2 \,\partial(p/T)/\partial T$ is the energy density. At $\mu_f=0$, the equation of state conditioning the pressure on the energy density could be determined with a high precision from first-principle lattice QCD calculation, see refs. \cite{Borsanyi:2016ksw, Aoki:2009sc, Bazavov:2014pvz, Karsch:2013fga, Bernard:2004je,Hands:2001ee, Kogut:2001na, Kogut:2001if,Kogut:2002kj,Aoki:2006we,Alles:2006ea,Hands:2006ve,Hands:2010gd,Bazavov:2011nk,Borsanyi:2011sw} and be utilized in describng various physical systems, such as, evolution of the early Universe \cite{Tawfik:2011sh,Tawfik:2019jsa}, relativistic heavy-ion collisions  \cite{Shen:2007zze} and stellar compact objects \cite{Kampfer:1983zz, Janka:2012wk, Takahara:1985qcm, Pons:2001ar, Sagert:2008ka, Fischer:2010zzb, Beisitzer:2014kea}.
    
The Stefan-Boltzmann (SB) limit can be determined from the grand canonical partition function of an ideal gas. In limit of infinite temperature, the thermodynamic pressure of $N_f$ quark flavors is given as \cite{Kapusta:2006book, Chatterjee:2012np}
\bea
\frac{p_{SB}}{T^4} &=& \frac{19\,\pi^2}{36} +\sum_{f} \left[ \frac{1}{2} \left(\frac{\mu_f}{T} \right)^2+\frac{1}{4\,\pi^2} \left(\frac{\mu_f}{T} \right)^4\right],
\label{SBlimit}
\eea 
where the first term refers to the contributions of quarks and gluons at vanishing chemical potential. The second term indicates contributions of ideal gas at finite chemical potential. We can utilize this expression to  determine  straightforwardly the ideal gas limit for bulk thermodynamic quantities, including susceptibilities and correlations \cite{Chatterjee:2012np}.
   
\begin{figure}[htb]
\centering{
\includegraphics[width=16cm,height= 11  cm,angle=0]{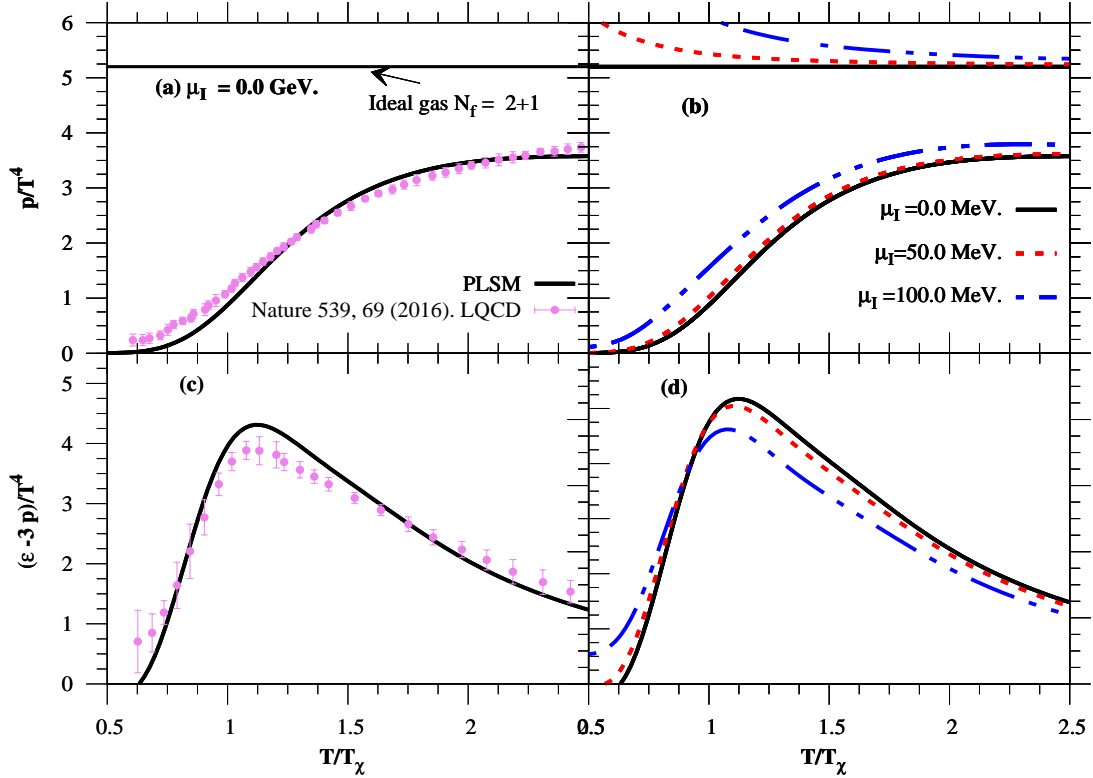} 
\caption{\footnotesize (Color online) Upper panel: left hand panel shows the temperature dependence of the normalized PLSM  $p/T^4$ (solid curve) compared with lattice QCD calculations \cite{Borsanyi:2016ksw} (closed symbols) at $\mu_I=0$. Right hand panel compares the results at $\mu_I=0$ with predictions at $\mu_I=50~$MeV (dotted) and $100.0~$MeV (dash-double-dotted curve). The $(2+1)$ SB-limit shown in upper part is related to the isospin chemical potential.  Lower panel:  The same as the upper panel but here for $(\epsilon-3p)/T^4$.
\label{Fig.pressure}}
}
\end{figure}

Upper panel of figure \ref{Fig.pressure} depicts $p/T^4$ as a function of $T$ at vanishing (left panel) and finite $\mu_I$ (right panel). At $\mu_I=0$, the PLSM $p/T^4$ is confronted to recent lattice QCD calculations \cite{Borsanyi:2016ksw}. No fitting was conducted. We merely compare results from both approaches. There is a convincing qualitative agreement. But, at low $T$, it seems that our calculations slightly underestimate the lattice QCD results. Same situation appears at large $T$. At temperatures around $T_\chi$, the agreement becomes relatively good. One has to bear in mind that the Polyakov-loop potential proposed likely plays an great role in the results obtained from PLSM including the thermodynamic quantities. A detailed discussion on the impacts of various Polyakov-loop potentials could be found in refs. \cite{Ratti:2005jh,Roessner:2006xn,Schaefer:2007pw,Fukushima:2008wg, Lo:2013hla}. It is worthy highlighting that the lattice QCD and the PLSM results at the highest temperature are about $32\%$ below the SB limit.

These results allow us to conclude that the phase transitions in both approaches have the same order; crossover. Also, the corresponding $T_\chi$ likely agree with each other. With this regard, it is worthy mentioning that the critical temperature is not universally constant even in the lattice QCD calculations \cite{Borsanyi:2016ksw}, which is mainly depending on various input parameters for the lattice QCD simulations. For the seek of a good comparison, the pseudo-critical temperature $T_\chi(\mu_f)$ in the present calculations, at vanishing baryon chemical potential, was approximately estimated as $T_\chi=210~$MeV. 

The right panel of Fig. \ref{Fig.pressure} (b) shows that same as in panel (a) but here the results at $\mu_I=0.0$ (slid curve) are compared with predictions at $\mu_I=50.0~$MeV (dotted Curve) and $\mu_I=100~$MeV (dash-double-dotted curve). We observe that with increasing $\mu_I$, the values of $p/T^4$ increase, as well, and move to the left-hand side, i.e. to lower $T$. To the authors' best knowledge, there are no lattice QCD calculations at finite $\mu_I$ possible to compare with.


The thermodynamic quantities $p$ and $\epsilon$ characterizing the EoS can also be expressed in terms of the interaction measure $\Delta=\epsilon -3p$ and speed of sound squared $c_s^2= \partial p/\partial \epsilon$. Lower panel of figure \ref{Fig.pressure} shows $\Delta/T^4$ in dependence on $T$. In the confined phase, $\Delta/T^4$ gradually increases with the increase in $T$. The phase transition, the smooth crossover, could be characterized, where $\Delta/T^4$ flips, i.e. becomes decreasing with the increase in $T$. But it should be noiced that the peak of $\Delta/T^4$ is related to the change of $\Delta/T^4$ rather than the phase transition, itself \cite{Mao:2009aq}. We find that there is a good agreement with the lattice QCD calculations \cite{Borsanyi:2016ksw}. 

Besides the interaction measure $\Delta = \epsilon -3p$, the behavior of the thermodynamic quantities $\epsilon$ and $p$ could be used to determine the order of the phase transition from confined to deconfined phases, could be used as thermodynamic order parameters. The QCD asymptotic freedom implies that the interaction measure becomes dependent on the strength of the  running strong  coupling as $\propto \alpha_s^2 T^4$ \cite{Tawfik:2013eua}. In terms of scale invariant theory, the interaction measure indicating the chiral phase structure, the trace anomaly, is assumed very small for freely colliding partons and for hadronic fluid. At $T<T_\chi$, the trace anomaly is apparently sufficiently small. This is characterized by an increasing interaction strength, which equivalently tends to bring quarks and gluons close to each others. This picture would illustrate the reason that quarks and antiquarks are bound forming hadrons. When $T>T_\chi$, it is apparent that $\alpha_s$ becomes small. Accordingly, the interaction strength becomes weaker and weaker. This means that the quarks and gluons form an ideal gas, especially at very high $T$, where $\Delta\approx 0$. 

The right panel of Fig. \ref{Fig.pressure}  (d) compares the PLSM results at $\mu_I=0.0$ (slid curve) with the results at $50.0$ (dotted Curve) and $100~$MeV (dash-double-dotted curve). We notice that the increase in $\mu_I$ tends to displace the results, i.e. moving these to lower temperatures. Also, we notice that increasing $\mu_I$ leads to smoothing the temperature dependence. It is obvious that  the pseudo-critical temperature $T_\chi$ decreases with increasing $\mu_I$, Fig. \ref{QCDdiagram}.

\subsection{Fluctuations of conserved quantum charges}
\label{Fluc}
The fluctuations plays an essential role in particle physics. For example, they are proposed as signatures for the chiral phase transition \cite{Borsanyi:2011sw, Bleicher:2000ek}. Various lattice QCD simulations aim at determining these quantities, see for instance refs. \cite{Borsanyi:2011bm, Cheng:2008zh, Chatterjee:2011jd, Chatterjee:2012np}. Moreover, various effective QCD-like models presented similar calculations \cite{Ghosh:2007wy,Borsanyi:2011sw}.  The fluctuations of different quantum charges, such as baryon $B$, strangeness $S$, electric charge $Q$, and isospin $I$ can be derived from the pressure with respect to the independent thermodynamic quantities $T$ and $\mu_f$. Equation (\ref{muf-eqs}) expresses $\mu_f$ of conserved quantum charges, which as well are considered as independent variables in grand canonical ensemble \cite{Bzdak:2012an}. The thermal expectation values of the conserved charges $X=[B, Q, I, S, \cdots]$, the extensive variables, can be derived from the derivative of the grand canonical partition function $\mathcal{Z}$ with respect to corresponding chemical potential $\mu_X$,
\bea
\left\langle N_X \right\rangle & = & T\, \frac{\partial \ln{ [\mathcal{Z}(V,\; T,\; \mu_f)]}}{\partial \mu_X}. 
\eea   
The second derivative of $\mathcal{Z}$, known as susceptibility, leads to,
\bea
\frac{\partial\; \left\langle N_X \right\rangle}{\partial \mu_Y} &=& T\, \frac{\partial^2 \ln{ [\mathcal{Z}(V,\; T,\; \mu_f)]}}{\partial \mu_Y \; \partial \mu_X} = \frac{\left\langle N_X \,N_Y\right\rangle -\left\langle N_X \right\rangle\;\left\langle N_Y \right\rangle}{T}.
\eea 
The fluctuations of two identical quantum numbers, i.e. $X=Y$, express correlations instead. The thermodynamic pressure can be related to the thermodynamic potential, Eq. (\ref{potential}) and  Eq. (\ref{ThermoEQs}). In light of this discussion, a generic expression for the fluctuations and the correlations can be obtained 
\bea
\chi_{ijkt}^{BQIS} =\frac{\partial^{i+j+k+l} \; p/T^4}{(\partial\;\hat{\mu}_B)^i\; (\partial\;\hat{\mu}_Q)^j\;(\partial\;\hat{\mu}_I)^k\;(\partial\;\hat{\mu}_S)^l\; }, \label{Fluctuations}
\eea
where $\hat{\mu}_X=\mu_X/T$. The normalization to $T^4$ is there to ensure that the cumulants remain dimensionless \cite{Borsanyi:2011sw}. As a mentioned above, the PLSM simulations for the second order fluctuations or susceptibilities, i.e. quadratic fluctuations where $i+j+k+l=2$. The PLSM results are compared with recent LQCD simulations.

\begin{figure}[htb]
\centering{
\includegraphics[width=13.cm, height=12.cm,angle=0]{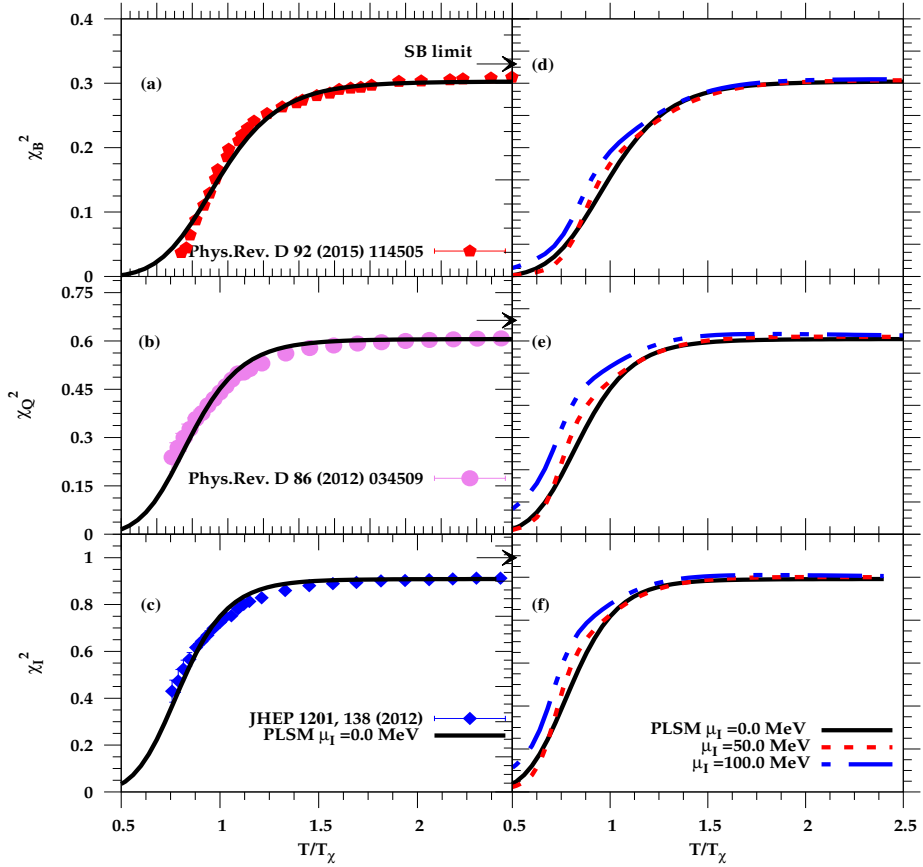}
\caption{\footnotesize (Color online) Left panel shows the net-baryon number (a), net-electric-charge (b) and net-isospin (c) fluctuations calculated in the PLSM as functions of $T$, at $\mu_I=0$. The PLSM results are compared with recent lattice QCD simulations (symbols) \cite{Borsanyi:2011sw, Bellwied:2015lba}. Right panel depicts the same as in left panel but here at $\mu_I=0.0$ (solid curve), $50.0$ (dotted curve), and $100.0~$MeV (dash-double-dotted curve). The SB-limits are drawn in the upper corners of the graphs.
\label{chargeFulL}}
}  
\end{figure}

Fig. \ref{chargeFulL} shows the temperature dependence of the net-baryon number (top), net-electric-charge (middle) and net-isospin fluctuations (bottom panel) as calculated from PLSM  as functions of $T$, at vanishing (left panel) and finite $\mu_I$ (right panel). Left panel focuses on the results at vanishing $\mu_I$ (solid curves). The net-baryon (a), electric-charge (b), and isospin fluctuations are compared to recent lattice QCD calculations (symbols) \cite{Borsanyi:2011sw, Bellwied:2015lba}. Fig. \ref{chargeFulL} points out that the susceptibilities seem to vanish at low temperatures. This can be understood from the observation that the chiral condensates have large values in this region of temperatures. Accordingly, the large quark masses of the relevant degrees of freedom associated with is responsible for small fluctuations. The increase in the temperature releases more degrees of freedom and accordingly small masses. With the degrees of freedom we mean all hadron states in the confined phase and quarks and gluons in the deconfined phase. The existence of a rapid change (increase) apparently indicates some kind of phase transition, see for instance refs. \cite{Borsanyi:2011bm, Cheng:2008zh, Chatterjee:2011jd, Chatterjee:2012np}. At high temperatures, where small quark masses are assigned to each of the effective degrees of freedom, the confined system is believed to form a new state-of-matter (deconfined massless quarks).

\begin{table}[htb]
\begin{center}
\begin{tabular}{|c|*{5}{c|}r} 
\hline
Cumulants in SB limit  & $B$ & $Q$ & $S$ &$I$  \\ 
\hline \hline  
  $\chi_2^{X}$  & $1/3$ & $2/3$ & $1$ & $1$\\ 
\hline 
\end{tabular}
\caption{The fluctuations for baryon number ($B$), electric charge ($Q$), strangeness ($S$) and isospin ($I$) in an ideal gas limit, SB limits.}  \label{tab:SB}
\end{center}
\end{table}

The corresponding SB-limit of ideal gas for $N_f=2+1$ are estimated as shown in Tab. \ref{tab:SB}. Accordingly, further reliable conclusions can be drawn. At low $T$,  the  conserved charge fluctuations are small, as well. With increasing $T$, the fluctuations increase. This continues until approaching a kind of stability. Again, the latter characterizes some kind of transition to the deconfined phase. To summarize, there are  small fluctuations at low $T$ as result of the large masses of the relevant degrees of freedom, as $T$ increases, a rapid increase in $T_\chi$ similar to the one related to the other thermodynamic quantities, such as pressure, Fig. \ref{Fig.pressure}, takes place. At high $T$, the effects of $\mu_I$ becomes negligibly small. However, the excellent agreement with the lattice QCD results obtained indicates that the PLSM with the parameters given in Tab. (\ref{tab:1a}), the type of the potential of Polyakov loops and the constants, well reproduce the lattice results. In light of this, solid conclusions could be drawn.

The right panel compares the same results at $\mu_I=0.0$ (solid curves) with the PLSM results, at $50.0$ (dotted), and $100.0~$MeV (dash-double-dotted curves). Unfortunately, there are no lattice QCD calculations at finite $\mu_I$ to compare with. Our PLSM results, which excellently reproduce the lattice results at vanishing $\mu_I$ are likely able to present reliable predictions, from which we observe that increasing $\mu_I$ slightly shifts the values of the fluctuations of the various quantum numbers to the left, i.e. to smaller temperatures. This apparently indicates that increasing $\mu_I$ slightly decreases $T_\chi$,  where the thermal behaviors of susceptibility are delayed as the isospin density increases, especially within the region of phase transition. 
 
 \begin{figure}[htb]
\centering{
\includegraphics[width=17.cm, height=6.5 cm,angle=0]{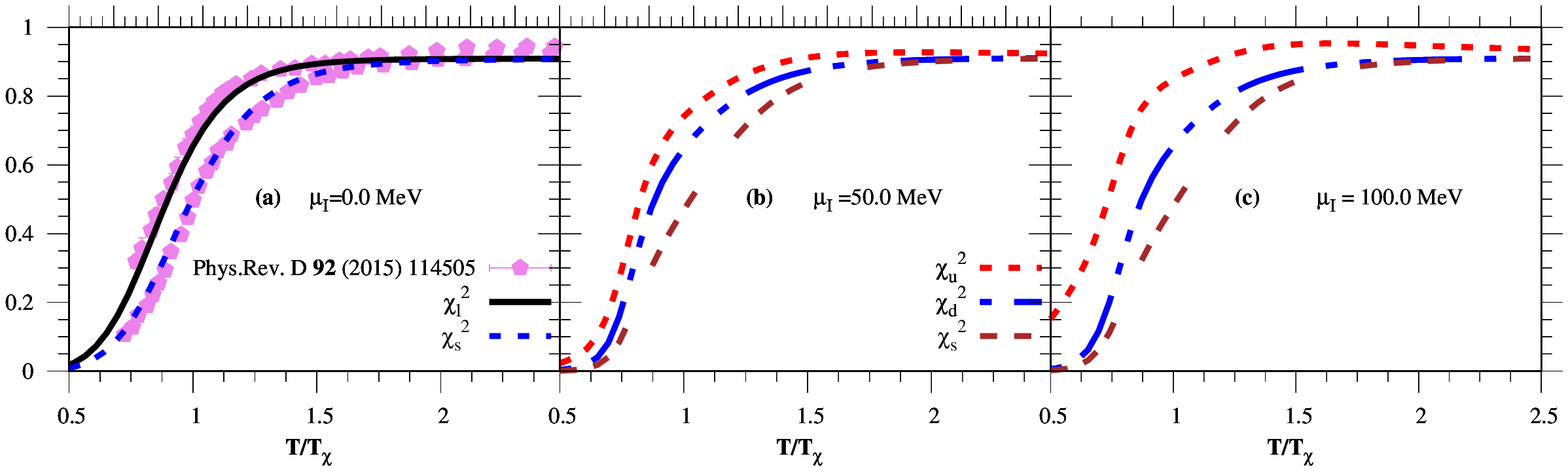}
\caption{\footnotesize (Color online) Left panel shows the susceptibilities of the $f^{th}$-quark flavor $\chi_f^2$ as functions of $T$ at $\mu_I=0.0$ compared with lattice QCD calculation \cite{Bellwied:2015lba}. The middle and right panels illustrate $\chi_f^2$ at $\mu_I=50.0$ and $100.0~$MeV, respectively.   
\label{qaurkFulL}}
}
\end{figure}

From Eq. (\ref{Fluctuations}) and by replacing the subscript $X$ by $f=[u,\,d,\,s]$, the fluctuations of $f^{th}$ quark flavors ($\chi_f^2$) can be calculated as functions of $T$ and $\mu_I$, Fig. \ref{qaurkFulL}.  The left panel gives the temperature dependence of the nonstrange $\chi_l^2$ (solid) and strange $\chi_s^2$ quark number fluctuations (dashed curve) at $\mu_I=0.0~$MeV and compares the PLSM results with lattice QCD calculations \cite{Bellwied:2015lba}. There is good agreement in both types of fluctuations. We observe that at $T<T_{\chi}$ the fluctuations increase with increasing $T$.  At $T>T_{\chi}$, the fluctuations become $T$ independent. As discussed in earlier sections, the masses associated with the quarks play an essential role in the fluctuations. At low $T$, the system is confined and likely has large masses. This restricts the fluctuations. As $T$ increases the system undergoes phase transition to deconfined phase. This process seems being not sudden or prompt. It takes place until the system is completely transformed to a massless quarks and gluons at high $T$, i.e. forming an ideal gas. 

At vanishing isospin asymmetry, the $u$- and $d$-quark susceptibilities are obviously not distinguishable. At finite isospin asymmetry, the middle and right panels, the PLSM results on $\chi_u^2$, $\chi_d^2$, and $\chi_s^2$ at $\mu_I=50.0$ and $100.0~$MeV, respectively, are represented by dotted, dash-double-dotted, and dashed curves, respectively. We observe that the $u$-quark susceptibilities have higher values than that of the $d$- and $s$-quarks, respectively. Again, the isospin asymmetry is assumed to distinguish between both components in the nonstrange quark sectors; up and down. At vanishing isospin asymmetry, an excellent agreement between the PLSM and lattice QCD results is obtained, left panel (a) of Fig. \ref{qaurkFulL}. 

In middle (b) and right (c) panels of Fig. \ref{qaurkFulL}, the nonstrange quark susceptibilities become  distinguishable, especially at finite isospin asymmetry. The temperature dependence of $u$-quark  susceptibility is apparently larger than that of $d$- and $s$-quark, respectively. This observation seems to support the conclusion that the heavy quarks have smaller fluctuations and vice versa. The pseudo-critical temperature $T_{\chi}$, whose estimation was elaborated in previous sections is also located within the deconfinement  phase transition. We observe that the region of the phase transition greatly increases, i.e. is shifted to higher temperatures, with increasing temperature and also with increasing isospin chemical potential. We also conclude that the pseudo-critical temperature is not an universal constant but it is strongly dependent on the quark content. Furthermore, the PLSM results seem to confirm that $T_{\chi}$ decreases with increasing $\mu_I$. Last but not least, we find that the effects of $\mu_I$ on the temperature dependence of $\chi_f^2$ seems negligible at higher $T$, where the quarks are conjectured deconfining and moving almost freely. 
\subsection{QCD phase diagram \label{QCD}}  

\begin{figure}[htb]
\centering{
\includegraphics[width=6.5cm,angle=-90]{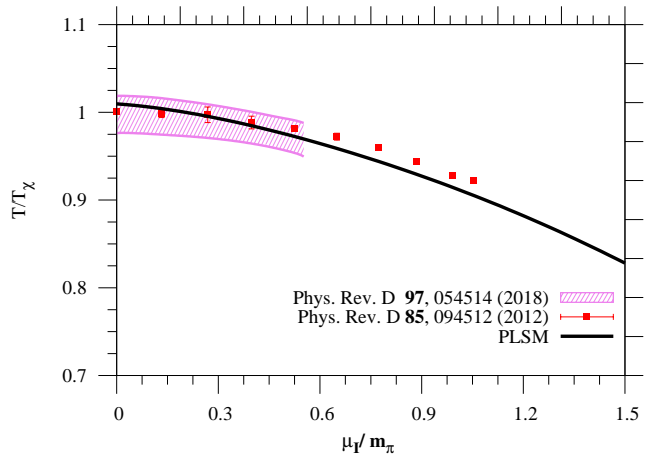}
\caption{\footnotesize (Color online) QCD phase diagram at vanishing baryon chemical potential (vanishing net quark density) but finite isospin chemical potential. The PLSM results (solid curves) are confronted to recent lattice QCD calculations (symbols) \cite{Brandt:2017oyy, Cea:2012ev}.  
\label{QCDdiagram}}
}
\end{figure}

In this section, we summarize the main results of this study. First, we conclude that the impacts of finite isospin asymmetry seem to enhance the various PLSM results, such as bulk thermodynamic quantities including susceptibilities and second-order fluctuations of various quantum charges with increasing temperature. Second, when mapping out the PLSM temperatures versus isospin asymmetry, an extension of the QCD phase structure to finite isospin chemical potential was achieved.  We find that the characteristic pseudo-critical temperature decreases as the isospin asymmetry increases. Third, when the PLSM results on the pseudo-critical temperatures are confronted to recent lattice QCD simulations \cite{Brandt:2017oyy, Cea:2012ev}, an excellent agreement is obtained, Fig. \ref{QCDdiagram}. For the seek of a reliable agreement, our results on temperature and isospin chemical potential are - similar to lattice QCD simulations - normalized to the pseudo-critical temperature and the pion mass, respectively. For lattice QCD, $m_\pi=400.0~$MeV and $T_\chi^{\mu_I=0}=164~$MeV were utilized, while for the PLSM results, we use $m_\pi=138~$MeV and $T_\chi^{\mu_I=0}=210~$MeV. 
       
From Fig. \ref{QCDdiagram}, we find that the pseudo-critical temperature decreases with the increase in $\mu_I$. This behavior is the same as that obtained in lattice QCD simulations \cite{Brandt:2017oyy,Cea:2012ev}. That the PLSM results well reproduce the available lattice QCD calculations, we can straightforwardly predict the tendence at larger $\mu_I$ exceeding the ones covered by the available lattice QCD calculations. When comparing Fig. \ref{QCDdiagram} with the ones reporting on the QCD phase structure in ($T_{\chi}$-$\mu_B$) plane \cite{Tawfik:2019rdd,Tawfik:2017cdx,Tawfik:2016gye,Tawfik:2014uka}, we realize that both planes look similar. Further studies combining the critical temperatures with both types of chemical potentials are planned in the near future.

%
\section{Conclusions} 
\label{conclusion}

We have studied the SU($3$) Ployakov linear-sigma model (PLSM) with U$(1)_A$ anomaly at finite isospin asymmetry, which enables us distinguishing between the chiral phase transitions corresponding to each of the light quark flavors; $\sigma_u$ and $\sigma_d$. In SU$(3)$, finite isospin asymmetry makes the mean sigma-fields $\bar{\sigma_a}$ having nonzero diagonal generators as $\bar{\sigma}_0 \ne \bar{\sigma}_3 \ne \bar{\sigma}_8 \ne 0 $  and the parameters of explicity symmetry breaking are nonvansihing $h_0 \ne h_3 \ne h_8 \ne 0 $. In other words, the impacts of finite $\sigma_3$ and $h_3$ break SU$(2)$ isospin asymmetry, where $\sigma_u=\sigma_l+\sigma_3$ and $\sigma_d=\sigma_l-\sigma_3$ \cite{Lenaghan:2000ey, Schaffner-Bielich:2012bda, Beisitzer:2014kea}. To this end, we first drive the thermodynamic potential of the pure mesonic contributions in SU$(3)$ in basis of quark flavors $\sigma_u$, $\sigma_d$ and $\sigma_s$. Second, we include in it Polyakov-loop potential in order to assure integrating the gluonic degrees of freedom in the chiral LSM and the gluon-quark interactions. 

We have estimated $\sigma_u$, $\sigma_d$, $\sigma_s$, $\phi$, and $\bar{\phi}$. These are evaluated by minimizing the real part of thermodynamic potential $\mathcal{R}e\;  [\Omega (T,\,\mu_f)]$, Eq. (\ref{potential}), at saddle point in order to estimate the gap-equations  of the PLSM, App. \ref{App_A}, as functions of two independent variables; temperature and quark chemical potential. We found that a common feature can be reported; for chiral phase transition the pseudo-critical temperatures decrease as the isospin chemical potential increases.

Various thermodynamic quantities including pressure and interaction measure have been analyzed.  Also, the fluctuations in form of second-order moments of different quantum numbers, such as baryon number, electric charge, and isospin calculated in PLSM were confronted to recent lattice QCD calculations.  Accordingly, the various parameters in PLSM could be fixed, reliably. All these analyses contributed to the characterization of QCD phase structure at finite isospin chemical potential. We observed that the pseudo-critical temperatures are not  universally constants but vary with quark flavors and apparently with the increase in the isospin chemical potentials. From second-order correlations and fluctuations, we found that increasing isospin chemical potential enhances and shifts these quantities. Last but not least, we conclude that the QCD phase structure in the ($T_\chi$-$\mu_I$) plane looks very similar to the one in the ($T_\chi$-$\mu_B$) plane.
 
\appendix 

\section{U($3$) algebra \label{App_algebra}}
The generator operator  $\hat{T}_a= \hat{\lambda}_a/2$ in U($3$) is a obtained from Gell-Mann matrices $\hat{\lambda}_a$ \cite{Weinberg:1972kfs} with the indices running as $a=0,\cdots,\,8$. From U($3$) algebra, we have 
\bea
\left[\hat{T}_a, \;\hat{T}_b \right] &=& i f_{abc} \hat{T}_c, \\
\left\{\hat{T}_a, \;\hat{T}_b\right\} &=& i d_{abc} \hat{T}_c, 
\eea
where $f_{abc}$ and $d_{abc}$ are the standard antisymmetric and symmetric structure constants of SU($3$), respectively.  The symmetric structure constant $d_{abc}$ can be defined as 
\bea
d_{abc} &=& \frac{1}{4} \mathit{Tr} \left[  \left\{ \hat{\lambda}_a,\; \hat{\lambda}_b \right\}  \hat{\lambda}_c \right], \\
d_{ab0} &=& \sqrt{\frac{2}{3}} \; \delta_{ab}.
\eea
In PCAC relation, the decay constant $f_a$ is related to the symmetric structure constant as  
\bea
f_a = d_{aab} \bar{\sigma}_a.
\eea  
Accordingly, the decay constants of the charged and neutral pion mesons ($f_{\pi^{\pm}} =f_1,\;f_{\pi^0} =f_3$) and kaon meson ($f_{K^{\pm}} =f_4,\;f_{K^0} =f_6$)  are given as 
\bea
f_{\pi^0} &=& f_{\pi^{\pm}} =\sqrt{\frac{2}{3}} \bar{\sigma}_0 + \frac{1}{\sqrt{3}} \bar{\sigma}_8, \\
f_{K^{\pm}} &=& \sqrt{\frac{2}{3}} \bar{\sigma}_0 + \frac{1}{2} \bar{\sigma}_3 -  \frac{1}{2\sqrt{3}}  \bar{\sigma}_8, \\
f_{K^{0}} &=& \sqrt{\frac{2}{3}} \bar{\sigma}_0 - \frac{1}{2} \bar{\sigma}_3 -  \frac{1}{2\sqrt{3}}  \bar{\sigma}_8,
\eea
where the isospin sigma field, $\bar{\sigma}_3$, is the difference between the decay constants of neutral and charged kaon mesons as,  
\bea
\bar{\sigma}_3 &=&  f_{K^{\pm}}-f_{K^{0}}.
\eea
From the experimental and recent lattice review on physical constants \cite{Barnett:1996hr, Tanabashi:2018oca,Aoki:2016frl},  $f_{\pi^{\pm}} = f_{\pi^0} =92.4\; \mathrm{MeV}$ and $f_{K^{\pm}} =113\; \mathrm{MeV} , \;f_{K^0} =113.453 \;\mathrm{MeV}$.

\section{Mean-field approximation \label{App_A}}

In order to perform the grand potential of PLSM in mean field approximation, we start from the partition function $\mathcal{Z}$. In thermal equilibrium, the exchanges of energies between quarks and antiquarks can be given by the path integral over all fermions and bosonss such as
\begin{eqnarray}
\mathcal{Z}&=& \mathrm{Tr \;exp}[-(\hat{\mathcal{H}}-\sum_{f=u, d, s} \mu_f \hat{\mathcal{N}}_f)/T] \nonumber\\
&=& \int\prod_a \mathcal{D} \sigma_a \mathcal{D} \pi_a \int
\mathcal{D}\psi \mathcal{D} \bar{\psi}\;   \mathrm{exp} \left[ \int d^4\;x
(\mathcal{L}+\sum_{f=u, d, s} \mu_f \bar{\psi}_f \gamma^0 \psi_f )
\right].
\end{eqnarray}
 
The various PLSM order parameters in mean-field approximation can be utilized in evaluating the temperature dependence of the meson sigma fields $\bar{\sigma_u},\;\bar{\sigma_d}\; \mbox{and}\; \bar{\sigma_s}$, and Polyakov-loop variables $\phi$ and $\bar{\bar{\phi}}$. We minimize the thermodynamic potential, Eq. (\ref{potential}), with respect to the expectation value of these parameters 
\bea
\left.\frac{\partial \Omega}{\partial \bar{\sigma_u}}=\frac{\partial \Omega}{\partial \bar{\sigma_d}}= \frac{\partial \Omega}{\partial \bar{\sigma_s}} = \frac{\partial \Omega}{\partial
\bar{\phi}} = \frac{\partial \Omega}{\partial \bar{\bar{\phi}}}\right|_{min}=0.  \label{Gap-eq}
\eea
The order parameters can be analyzed by minimizing the real term of the thermodynamic potential ($\mathcal{R}e\; \Omega$) at saddle point. In doing of this, we can estimate the behaviors of the expectation value of the chiral condensates and the Polyakov-loop variables as functions of two independent variables; temperature  and chemical potential.


\bibliographystyle{aip}

\bibliography{Atawfik_FiniteIsospin_IJMPA1} 

\end{document}